\documentclass[a4paper, 12pt]{article}

\usepackage{graphicx}
\usepackage{url}
\usepackage{amsbsy,amssymb} 
\usepackage{amsmath}
\usepackage{abstract}
\usepackage{hyperref}
\usepackage{color}

\usepackage{lscape}

\newtheorem{Proof}{Proof}
\newtheorem{Lemma}{Lemma}

\begin{document}

{\LARGE\centering{\bf{Integrability of geodesics of totally geodesic metrics}}}

\begin{center}
\sf{Rados\l{}aw A. Kycia$^{1,2,a}$, Maria U\l{}an$^{3,b}$}
\end{center}

\medskip
\small{

\centerline{$^{1}$The Faculty of Science, Masaryk University}
\centerline{Kotl\'{a}\v{r}sk\'{a} 2, 602 00 Brno}
\centerline{Czechia}

\centerline{$^{2}$Cracow University of Technology, Faculty of Physics}
\centerline{Mathematics and Computer Science, PL-31155, Krak\'{o}w}
\centerline{Poland}

\centerline{$^{3}$Baltic Institute of Mathematics}
\centerline{Wa\l{}brzyska 11/85, 02-739 Warszawa}
\centerline{Poland}

\centerline{$^{a}${\tt
kycia.radoslaw@gmail.com}}

\centerline{$^{b}${\tt
maria.ulan@baltinmat.eu}}

}

\begin{abstract}
\noindent
Analysis of the geodesics in the space of signature $(1,3)$ that splits in two-dimensional distributions resulting from the Weyl tensor eignespaces - hyperbolic and elliptic ones - described in [V. Lychagin, V. Yumaguzhin, \emph{Differential invariants and exact solutions of the Einstein equations}, Anal.Math.Phys. 1664-235X 1-9 (2016)] are presented. Cases when geodesic equations are integrable are identified. Similar analysis is performed for the same model coupled to Electromagnetism described in [V. Lychagin, V. Yumaguzhi, \emph{Differential invariants and exact solutions of the Einstein–Maxwell equation}, Anal.Math.Phys. 1, 19--29, (2017)].
\end{abstract}
%
PACS 2010: 04.20.Jb, 02.30.Ik, 04.40.Nr

\section{Introduction}
In \cite{SolutionsMain} a class of totally geodesic metrics were given. For convenience we cite here main steps referring interested reader to the paper for details.

The main point to start is to decompose the Weyl tensor in the base of 2-forms which are eigenvectors of corresponding Weyl operator. Then it results that the spacetime contains totally geodesic distributions \cite{SolutionsMain} of hyperbolic ($H$) and elliptic ($E$) tangent planes. This induces the solution of the Einstein's equations with cosmological constant $\Lambda$ in the form
\begin{equation}
 \begin{array}{c}
  g = g^{H} \oplus g^{E}, \\
  g^{H} = e^{\alpha(x_{0},x_{1})}(dx_{0}^{2}-dx_{1}^{2}), \quad g^{E} = -e^{\beta(x_{2},x_{3})}(dx_{2}^{2}+dx_{3}^{2}).
 \end{array}
 \label{MetricDecomposition}
\end{equation}

The functions $\alpha$ and $\beta$ are solutions of hyperbolic and elliptic Liouville equations, correspondingly, \cite{LiouvilleGeneral}
\begin{equation}
\left\{
 \begin{array}{c}
  \frac{\partial^{2}\alpha(x_{0},x_{1})}{\partial^{2}x_{0}} - \frac{\partial^{2}\alpha(x_{0},x_{1})}{\partial^{2}x_{1}}  + 2\Lambda e^{\alpha(x_{0},x_{1})} = 0, \\
  \frac{\partial^{2}\beta(x_{2},x_{3})}{\partial^{2}x_{2}} + \frac{\partial^{2}\beta(x_{2},x_{3})}{\partial^{2}x_{3}}  - 2\Lambda e^{\beta(x_{2},x_{3})} = 0.
 \end{array}
 \right.
 \label{LiouvilleEquations}
\end{equation}

The solution is as follows 
\begin{equation}
 \begin{array}{c}
  \alpha(x_{0},x_{1})=ln(h_{1}(v)(v_{x_{0}}^{2}-v_{x_{1}}^{2})),\\
  \beta(x_{2},x_{3})=ln(h_{2}(u)(u_{x_{2}}^{2}+u_{x_{3}}^{2})),\\
 \end{array}
\label{ansatz}
 \end{equation}
where $u$ and $v$ are solutions two-dimensional hyperbolic and elliptic equations:
\begin{equation}
 \begin{array}{c}
  v_{x_{0}x_{0}}-v_{x_{1}x_{1}}=0,\\
  u_{x_{2}x_{2}}+u_{x_{3}x_{3}}=0,
 \end{array}
\end{equation}
and where $h_{1}$ and $h_{2}$ are solutions of second order ODEs. Full list of the solutions is presented \cite{SolutionsMain}.

In this paper we analyse geodesic governed by (\ref{MetricDecomposition}). Calculations of geodesic equations were performed using Mathematica package CCGRG, see\cite{ccgrg1, ccgrg2, ccgrg3}, and symmetries using \emph{Differential Geometry} Maple package.

This paper is organized as follows: In the next section is is shown that there is no true singularities of geodesics in the model of \cite{SolutionsMain}, i.e. the spacetime is totally geodesic. Then the analysis of Liouville integrability \cite{Arnold} of the geodesics equations is provided. Finally, the analogous model with additional coupling to the Electromagnetic field described in \cite{EinsteinMaxwell} is considered in the terms of integrability of geodesics.

The presentation starts from the analysis of singularities of geodesics. 

\section{Singularities}

Metrics described in \cite{SolutionsMain} have obvious singularities. Generally, singularities in General Relativity have two origins \cite{Wald}:
\begin{itemize}
 \item {singularities of coordinates which results from the fact that in the coordinate patch ill-defined coordinate functions are used over regular points of manifold;}
 \item {true singularities which give geodesic incompleteness of the manifold;}
\end{itemize}

True singularities are usually visible in some invariants of curvature. The simplest second-order one is the square of the Riemann curvature (called Kretschmann scalar \cite{SecondOrderInvariants})
\begin{equation}
 K=R_{abcd}R^{abcd}.
 \label{KretschmannScalar}
\end{equation}
For (\ref{MetricDecomposition}) that are solutions of (\ref{LiouvilleEquations}) the invariant is constant
\begin{equation}
K= 8 \Lambda^{2},                                                                                                                                                                                           \end{equation}
which suggests no singularities, i.e., completeness of the pseudoriemannian manifold. The answer is affirmative as it is provided by the following Lemma\footnote{RK would like to thanks Igor Khavkine for discussion on this subject and suggestions of the outline of the proof.}

\begin{Lemma}
 The pseudoriemannian manifold (\ref{MetricDecomposition}) with (\ref{LiouvilleEquations}) is complete.
\end{Lemma}

\begin{Proof}
 From the metric decomposition (\ref{MetricDecomposition}) and the fact that
 \begin{equation}
  R^{H} = \sum_{i,j=0}^{1}R^{ij}_{..ij} = 2\Lambda, \quad R^{E} = \sum_{i,j=2}^{3}R^{ij}_{..ij}=2\Lambda,
 \end{equation}
The space factorizes into two-dimensional subspaces of constant curvature. These subspaces are isometric to spaces with no singularities according to the well known the Killing-Hopf theorem (see, e.g.  Theorem 6.3 in \cite{KillingHopf}). 
\end{Proof}
The theorem states that any singularity of (\ref{MetricDecomposition}), (\ref{LiouvilleEquations}) is artificial singularity only and can be removed by suitable change of coordinates.

\section{Geodesics}
In this section the analysis of geodesic equations will be provided. In the first part the canonical form of the geodesic equations and their symmetries will be presented. Then the (Liouville) integrable cases will be singled out.

\subsection{Geodesic equations}
As tangent space decomposes into two-dimensional subspaces therefore the geodesic equations consists two pairs consisting two coupled ODEs  for $\gamma(s)=(x_{0}(s), x_{1}(s), x_{2}(s), x_{3}(s))$, namely,
\begin{equation}
 \left\{
 \begin{array}{c}
  x_{0}''+x_{0}'x_{1}'\alpha_{x_{1}}+\frac{1}{2}\left( x_{0}' \right)^{2} \alpha_{x_{0}}+\frac{1}{2}\left( x_{1}' \right)^{2} \alpha_{x_{0}}=0 \\
  x_{1}''+ \frac{1}{2} \left( x_{0}' \right)^{2} \alpha_{x_{1}} + \frac{1}{2} \left( x_{1}' \right)^{2}\alpha_{x_1} + x_{0}'x_{1}' \alpha_{x_0} =0,
 \end{array}
 \right.  
 \label{Eq.Geodesic01}
\end{equation}
 \begin{equation}
\left\{
 \begin{array}{c}
  x_{2}''+x_{2}'x_{3}'\beta_{x_{3}}+\frac{1}{2}\left( x_{2}' \right)^{2}\beta_{x_2}-\frac{1}{2}\left( x_{3}' \right)^{2}\beta_{x_{2}} = 0 \\
  x_{3}''-\frac{1}{2}\left(x_{2}'\right)^{2}\beta_{x_3}+\frac{1}{2}\left( x_{3}' \right)^{2} \beta_{x_{3}}+x_{2}'x_{3}'\beta_{x_2}=0.
 \end{array}
\right.
\label{Eq.Geodesic23}
\end{equation}

These equations can be significantly simplified. Adding and subtracting equations (\ref{Eq.Geodesic01}) and then introducing the light-cone variables (characteristics of the wave equation): $x_{0}=\frac{z_{0}+z_{1}}{2}$ and $x_{1}=\frac{z_{0}-z_{1}}{2}$ one gets
\begin{equation}
\Delta_{1}(z_{0},z_{1}):\left\{
\begin{array}{c}
 z_{0}''+\frac{\partial \alpha(z_{0},z_{1})}{\partial z_{0}} \left( z_{0}' \right)^{2}=0 \\
 z_{1}''+\frac{\partial \alpha(z_{0},z_{1})}{\partial z_{1}} \left( z_{1}' \right)^{2}=0.
\end{array}
\right.
\label{Eq.Geodesic01LightCone}
\end{equation}

Symmetries of (\ref{Eq.Geodesic01LightCone}) can be found by assuming that the generator of symmetry is of the form: $X=f(s,z_{0},z_{1})\partial_{s} + g(s,z_{0},z_{1})\partial_{z_{0}}+h(s,z_{0},z_{1})\partial_{z_{1}}$ and solving the following system of PDEs
\begin{equation}
 \pounds_{X^{(2)}} \Delta_{1}(z_{0},z_{1})|_{\Delta_{1}(z_{0},z_{1})}=0,
\end{equation}
where $\pounds$ is the Lie derivative along $X^{(2)}$ - the second prolongation of $X$ to the jet space \cite{Vinogradov, Lychagin, Olver1, Olver2}. The result is
\begin{equation}
 X_{1}=(As+B)\partial_{s},
\end{equation}
where $A$ and $B$ are constants. This gives scaling and translation of $s$ variable, and results from the fact that (\ref{Eq.Geodesic01}) does not depends explicitly on $s$. The symmetry reflects the fact that the geodesics should not depend on re-parametrization in $s$ and is also connected with the fact that geodesic equations are variational and should posses such symmetries.

The same procedure can be applied to the second system of (\ref{Eq.Geodesic23}). In this case we have positively defined ('elliptic') metric, which suggest complex characteristics. It is therefore more appropriate to use complex-valued characteristics of elliptic equation, i.e., the substitution $x_{2}=\frac{z_{2}+z_{3}}{2i}$ and $x_{3}=\frac{z_{2}-z_{3}}{2}$, where $i=\sqrt{-1}$. Then adding and subtracting from the first equation of (\ref{Eq.Geodesic23}) multiplied by the imaginary unity the second one  one gets the system which resembles (\ref{Eq.Geodesic01LightCone}), namely,

\begin{equation}
\Delta_{1}(z_{2},z_{3}):\left\{
\begin{array}{c}
 z_{2}''+\frac{\partial \beta(z_{2},z_{3})}{\partial z_{2}} \left( z_{2}' \right)^{2}=0 \\
 z_{3}''+\frac{\partial \alpha(z_{2},z_{3})}{\partial z_{3}} \left( z_{3}' \right)^{2}=0.
\end{array}
\right.
\label{Eq.Geodesic23LightCone2}
\end{equation}

Since the equations are the same as in the previous case, symmetry analysis indicates, as above, the following generator
\begin{equation}
 X_{2}=(Cs+D)\partial_{s},
\end{equation}
where $C$ and $D$ are some constants.

In the next section integrability of geodesics equations will be investigated.

\subsection{Integrability of geodesic equations}

First, let us consider hyperbolic part of the metric, namely define hamiltonian
\begin{equation}
 H_{0,\alpha}=e^{\alpha(x_{0},x_{1})}(p_{0}^{2}-p_{1}^{2}),
\end{equation}
which surfaces of constant value determine the movement of the particles (positive - massive particles, zero - massless particles). Since the submanifold dimension is $2$, therefore in order to find its foliation, according the the Liouville theorem \cite{Arnold}, one additional function that the Poisson bracket with $H_{0,\alpha}$ vanishes,  is needed. It is assumed in the polynomial form in $p_{0}$ and $p_{1}$, namely,
\begin{equation}
 H_{1,\alpha}=\sum_{k=0}^{n} f_{i}(x_{0},x_{1}) p_{0}^{k}p_{1}^{n-k},
\end{equation}
where $n$ is natural number that is fixed degree. Complete integrability is equivalent to the existence of the solutions of 
\begin{equation}
 \{H_{0,\alpha},H_{1,\alpha}\}_{PB}=0,
 \label{Eq.PoissonBracket0}
\end{equation}
where $\{.,.\}_{PB}$ is the standard Poisson bracket. Equation (\ref{Eq.PoissonBracket0}) gives the set of PDEs\footnote{All calculations for this section are available in the Maple files from: \url{https://github.com/rkycia/GeodesicsIntegrability}}. In order to check closeness of this system the Kruglikov-Lychagin multibracket \cite{KruglikovLychaginBracket1, KruglikovLychaginBracket2, KruglikovLychaginBracket3, KruglikovLychaginBracket4, KruglikovLychaginBracket5URL} is used. When applied on the system (\ref{Eq.PoissonBracket0}), it gives compatibility condition in terms of PDEs for $\alpha(x,y)$, which solution up to $n=5$ are
\begin{enumerate}
 \item {$n=1,2$: 
 \begin{equation}
 \alpha(x_{0},x_{1})= F\tanh(B(y-x)+A)^3+E\tanh(B(y-x)+A)^2+D\tanh(B(y-x)+A)+C;  
 \label{Eq.alpha_1}
 \end{equation}
 where $A,B,C,D,E,F$ are constants of integration and parametrize $\alpha$.
}
 \item {$n=3,4$: 
 \begin{equation}
  \alpha(x_{0},x_{1})= Ax+By+C;
  \label{Eq.alpha_2}
 \end{equation}
 where $A,B,C$ are constants of integration and parametrize $\alpha$.
} 
\end{enumerate}
Surprisingly, these solutions fulfil the first equation of (\ref{LiouvilleEquations}) only when the cosmological constant $\Lambda=0$. This is very prominent example of the role of the cosmological constant in integrability of geodesic equations.

For the case (\ref{Eq.alpha_1}) the integration can be easily performed using (\ref{Eq.Geodesic01LightCone}), and gives
\begin{equation}
\left\{
 \begin{array}{c}
  z_{0}(s)=As+J, \\
  \int_{0}^{z_{1}(s)} \exp(F\tanh(Ba+A)^3+E\tanh(Ba+A)^2+D\tanh(Ba+A)+C) da + Gs+H = 0,\\
 \end{array}
 \right.
 \label{Eg.SolAlpha_1}
\end{equation}
where the second solution is expressed in the implicit form, and $A,B,\ldots,F$ is as in (\ref{Eq.alpha_1}) and $G,H,J$ are constants dependent on initial data.

The second case (\ref{Eq.alpha_2}) can be explicitly expressed in terms of elementary functions, namely,
\begin{equation}
 \begin{array}{c}
  z_{0}(s)=-2\frac{\ln\left(\frac{2}{(Ds+E)(A+B)}\right)}{A+B}\\
  z_{1}(s)=-2\frac{ln\left(\frac{2}{(Fs+G)(A+B)}\right)}{A+B},
 \end{array}
 \label{Eg.SolAlpha_2}
\end{equation}
where $D,E,F,G$ are constants depending on initial data.

Similar analysis performed for the elliptic part of the metric, taking 
\begin{equation}
 H_{0,\beta}=e^{\beta(x_{2},x_{3})}(p_{2}^{2}+p_{3}^{2}),
\end{equation}
and 
\begin{equation}
  H_{1,\beta}=\sum_{k=0}^{n} f_{i}(x_{2},x_{3}) p_{2}^{k}p_{3}^{n-k},
\end{equation}
and checking when
\begin{equation}
 \{H_{0,\beta},H_{1,\beta}\}_{PB}=0,
 \label{Eq.PoissonBracket1}
\end{equation}
the two solutions for $\beta$ are obtained up to degree $n=5$, namely:

\begin{enumerate}
 \item {$n=1,2$:
 \begin{equation}
  \beta(x_{0},x_{1})= F\tanh(B(y-xi)+A)^3+E\tanh(B(y-xi)+A)^2+D\tanh(B(y-xi)+A)+C;  
  \label{Eq.beta_1}
 \end{equation}
where $A,B,C,D,E,F$ are constants of integration and parametrize $\beta$, and $i$ is imaginary unit.
 }
 \item {$n=3,4$:
 \begin{equation}
  \beta(x_{0},x_{1})= Ax+By+C;
  \label{Eq.beta_2}
 \end{equation}
 where $A,B,C$ are constants of integration and parametrize $\beta$.
 } 
\end{enumerate}
As in the previous case, these $\beta$'s solve (\ref{LiouvilleEquations}) only when the cosmological constant $\Lambda=0$.

For (\ref{Eq.beta_1}) the solution of (\ref{Eq.Geodesic23LightCone2}) is
\begin{equation}
\left\{
 \begin{array}{c}
  z_{2}(s)=As+J, \\
  \int_{0}^{z_{3}(s)} \exp(F\tanh(Ba+A)^3+E\tanh(Ba+A)^2+D\tanh(Ba+A)+C) da + Gs+H = 0,\\
 \end{array}
 \right.
 \label{Eg.SolBeta_1}
\end{equation}
where as before $G,H,J$ are integration constants depending on initial data.

For (\ref{Eq.beta_2}) the solution of (\ref{Eq.Geodesic23LightCone2}) is
\begin{equation}
 \begin{array}{c}
  z_{2}(s)=-2\frac{\ln\left(\frac{2}{(A-iB)(Ds+E)}\right)}{A-iB} \\
  z_{3}(s)=-2\frac{\ln\left(\frac{2}{(A+iB)(Fs+G)}\right)}{A+iB},
 \end{array}
 \label{Eg.SolBeta_2}
\end{equation}
where $D,E,F,G$ are again constants depending on initial data. These solutions are complex-valued, however, since $x_{2}$ and $x_{3}$ fulfilling real equations for geodesic therefore transforming to the original variables one gets real solutions.

In general geodesic equations can be composed selecting solution (\ref{Eg.SolAlpha_1}) or (\ref{Eg.SolAlpha_2}) for the hyperbolic part of the subspace, and (\ref{Eg.SolBeta_1}) or (\ref{Eg.SolBeta_2}) for the elliptic subspace. Therefore in total $4=2\times 2$ integrable solutions were obtained.

\section{Einstein-Maxwell solutions}
The results from the previous section can be used for analysis of the geodesics of the solutions for coupled Einstein and Maxwell equations described in \cite{EinsteinMaxwell}. In this model the totally geodesic solutions, the same as (\ref{MetricDecomposition}) solution, for the metric were obtained. However now $\alpha$ and $\beta$ are solutions of 

\begin{equation}
\left\{
 \begin{array}{c}
  \frac{\partial^{2}\alpha(x_{0},x_{1})}{\partial^{2}x_{0}} - \frac{\partial^{2}\alpha(x_{0},x_{1})}{\partial^{2}x_{1}}  + k_{1} e^{\alpha(x_{0},x_{1})} = 0, \\
  \frac{\partial^{2}\beta(x_{2},x_{3})}{\partial^{2}x_{2}} + \frac{\partial^{2}\beta(x_{2},x_{3})}{\partial^{2}x_{3}}  + k_{2} e^{\beta(x_{2},x_{3})} = 0,
 \end{array}
 \right.
 \label{LiouvilleEquationsEinsteinMaxwell}
\end{equation}
where 
\begin{equation}
 k_{1}=2\left( \frac{kJ}{c^{4}}+\Lambda \right), \quad k_{2}=\left( \frac{kJ}{c^{4}}-\Lambda \right).
\end{equation}
$\Lambda$ is cosmological constant, $k$ the gravitational constant, and the new parameter $J$ is connected with the solution for the Faraday tensor of electromagnetic field
\begin{equation}
 F=-2le^{\alpha(x_{0},x_{1})} dx_{0}\wedge dx_{1} + 2m e^{\beta(x_{2},x_{3})}dx_{2}\wedge dx_{3},
\end{equation}
where 
\begin{equation}
 l^{2}=\frac{J-I_{1}}{2}, \quad m^{2} = \frac{J+I_{1}}{2},
 \label{Eq.l2_m2}
\end{equation}
and where $I_{1}$ is invariant of characteristic polynomial of the skew symmetric operator $\hat{F}$ (associated to $F$ by $g(\hat{F}X,Y)=F(X,Y)$), namely its determinant. The $\pm l$ and  $\pm i m$, where $l,m \in \mathbb{R}$, are eingenvalues of hyperbolic and elliptic parts of the operator $\hat{F}$. 

The straightforward result from (\ref{Eq.l2_m2}) is that
\begin{equation}
 J=l^{2}+m^{2}, \quad I_{1}=m^{2}-l^{2}.
 \label{Eq.J_I1}
\end{equation}

From our previous considerations the metric is (Liouville) integrable when $k_{1}=0=k_{2}$, i.e., when $J=0$ and $\Lambda=0$. And therefore, since $l,m \in \mathbb{R}$, from the first equation (\ref{Eq.J_I1}) it results $l=0$ and $m=0$ and therefore the Faraday tensor vanishes. This shows that the integrable solutions for geodesics exist for no electromagnetic field and no cosmological constant in this model. The solutions for geodesics are exactly the same as in the previous section for the Einstein equations only, since the electromagnetic field vanishes.

\section{Discussion}
The semiriemmanian metric of (\cite{SolutionsMain}) describes anisotropic spacetime with distinction of some preferred space axis $x_{1}$ and therefore cannot describe the observed today spacetime where assumption on spherical symmetry is imposed. This distinguished space direction resembles the phenomena from phase transitions in solid state physics and therefore it suggest that the model can be applied in some phenomena that occurs when the universe undertake some kind of phase transition, e.g., in the early state of the universe. Similar description also applies to the coupled Einstein-Maxwell system.

Intriguing correspondence between vanishing cosmological constant and integrability of geodesic equations was noted. In case of electromagnetism for integrability also electromagnetic field must vanish.

\section{Conclusions}
In this paper analysis of geodesic of solution of the Einstein vacuum equations resulting from the Weyl tensor bivector structure was provided. In particular, integrable geodesic equations of special solutions of Einstein vacuum equation were found and described. Similar analysis was also performed for the Einstein-Maxwell system.

\section*{Acknowledgments}
We would like thanks prof. Valentin Lychagin and Igor Khavkine for enlightening discussions. We would also like thanks Sergey Tychkov for help to master Maple.
RK participation was supported by the GACR Grant 17-19437S, and MUNI/A/1138/2017 Grant of Masaryk University.




\end{document}